# Nonlinear thermovoltage and thermocurrent in quantum dots

S Fahlvik Svensson[1,7,8], E A Hoffmann[2,8], N Nakpathomkun[3], P M Wu[1], H Q Xu[1,4], H A Nilsson[1], D Sánchez[5], V Kashcheyevs[6] and H Linke[1,7]

[1] Solid State Physics and Nanometer Structure Consortium (nmC@LU), Lund University, Box 118, 221 00 Lund, Sweden

[2] Institute for Advanced Study, Technische Universität München, Lichtenbergstrasse 2a, D-85748 Garching, Germany

[3] Department of Physics, Thammasat University, Pathum Thani, Thailand, 12120

[4] Key Laboratory for the Physics and Chemistry of Nanodevices and Department of Electronics, Peking University, Beijing 100871, China

[5] Institut de Física Interdisciplinària i de Sistemes Complexos IFISC (UIB-CSIC), E-07122 Palma de Mallorca, Spain

[6] Faculty of Physics and Mathematics, University of Latvia, Riga LV-1002, Latvia

E-mail: sofia.fahlvik_svensson@ftf.lth.se, heiner.linke@ftf.lth.se

## Abstract

Quantum dots are model systems for quantum thermoelectric behavior because of the ability to control and measure the effects of electron-energy filtering and quantum confinement on thermoelectric properties. Interestingly, nonlinear thermoelectric properties of such small systems can modify the efficiency of thermoelectric power conversion. Using quantum dots embedded in semiconductor nanowires, we measure thermovoltage and thermocurrent that are strongly nonlinear in the applied thermal bias. We show that most of the observed nonlinear effects can be understood in terms of a renormalization of the quantum-dot energy levels as a function of applied thermal bias and provide a theoretical model of the nonlinear thermovoltage taking renormalization into account. Furthermore, we propose a theory that explains a possible source of the observed, pronounced renormalization effect by the melting of Kondo correlations in the mixed-valence regime. The ability to control nonlinear thermoelectric behavior expands the range in which quantum thermoelectric effects may be used for efficient energy conversion.

## 1. Introduction

A key driving force for exploring the thermoelectric (TE) properties of low-dimensional and mesoscopic devices is the increase in the TE performance that may be achieved by introducing sharp features in a device's electron transmission spectrum $\tau(E)$, where $E$ is the electron energy [1-4]. This expected performance enhancement is due to a combination of an electron-energy filtering effect that allows the generation of a large open-circuit voltage, $V_{th}$, in response to an applied temperature differential $\Delta T$ and the possibility to reduce the electronic thermal conductivity through violation of the Wiedemann-Franz law [5-7].

The same sharp features in $\tau(E)$ also make it important to consider nonlinear TE effects, for example, a $V_{th}$ that does not scale linearly with $\Delta T$. It is well established that a strongly varying $\tau(E)$ makes it very easy to drive a small device into nonlinear response (a breakdown of Ohm's law) using a purely electric driving force [8-10], making it reasonable to expect nonlinear effects when carriers are driven thermally. More generally, the use of TE devices is of practical interest primarily for large $\Delta T$, when the Carnot efficiency is large, that is, well outside the linear response regime ($\Delta T \ll T$) with respect to a thermal driving force.

---

[7] Authors to whom any correspondence should be addressed.
[8] These authors contributed equally to this work.

Quantum dots (QDs) are an attractive model system for fundamental studies of TE energy conversion [4, 11-16], because their $\tau(E)$, can be tuned. Moreover, the effect of $\tau(E)$ on the efficiency of TE energy conversion and on $V_{th}$ can be characterized in detail [17, 18]. In particular, QDs with a highly-modulated $\tau(E)$ can operate in principle with TE energy conversion near Carnot efficiency [3, 4] and the Curzon-Ahlborn limit [3, 4, 13, 14]. It is therefore of interest to determine: does an efficiency-enhancing, fast-varying $\tau(E)$ induce nonlinear TE behavior in QDs? From a theoretical point of view, a nonlinear $V_{th}$ in QDs has been predicted, for example, due to changes of the dot's energy spectrum as a function of $V_{th}$ [19], in the Kondo regime [20, 21], and for molecular junctions [22]. Very recently, general scattering theories for the weakly nonlinear TE regime in mesoscopic devices have been developed, taking into account self-consistent screening effects [23-25], which in general lead to transmission functions that depend both on the electrochemical potentials (the applied voltage) and the temperatures in the leads. For the specific case of a double-barrier resonant tunneling structure (a QD), a nonlinear thermovoltage [23] and rectification effects [23, 24, 26] have been predicted which, intriguingly, imply the possibility to enhance the TE efficiency beyond that expected in linear response when the effect of capacitive coupling to gates is considered [24, 27]. In experiments, a thermovoltage that changes nonlinearly with $\Delta T$ has been observed in semiconductor QDs [17, 28, 29] and in molecular junctions [30], but to date such nonlinear effects have not been explored systematically in experiments.

Here, we present an experimental study of nonlinear TE behavior in QDs formed by a double-barrier structure in semiconductor nanowires. With increasing thermal bias, and at constant electric bias, we observe strongly nonlinear TE signals, including sign reversals of $V_{th}$ [17] and the thermally induced electric current (thermocurrent), $I_{th}$. To understand our observations, we first consider the energy-dependence of the dot's measured electron transmission spectrum $\tau(E, V)$. Our analysis concludes that the variation of transmission probability with energy and voltage, $\tau(E, V)$, by itself can explain only part of the observed nonlinear behavior. We then measure the QD's energy spectrum as a function of applied $\Delta T$ and observe that the nonlinear effects appear to be primarily due to temperature-induced renormalization effects, that is, a shift of energy levels in response to $\Delta T$, resulting in a transmission function $\tau(E, \Delta T)$. To understand our observations, we consider a simple sequential tunneling model and show that the strongly nonlinear dependence indeed is caused by an energy-level renormalization originating from the QD response to the applied thermal bias [23]. Furthermore, we argue that an observed sharp thermally-driven change of the effective level position in one of our samples may come from melting of residual Kondo correlations in the mixed-valence regime.

## 2. Devices and experimental details

TE measurements were performed on three different nanowire-based QD devices in different material systems, demonstrating that the occurrence of nonlinear effects in QDs appears to be generic and independent of device details. The first QD, referred to as QD1, was defined by two InP barriers in a 66 nm diameter InAs nanowire grown by chemical beam epitaxy (CBE) [31]. The dot itself consisted of $InAs_{0.8}P_{0.2}$ and had a length of approximately 190 nm (all lengths and diameters were determined from scanning electron microscope (SEM) images after the measurements were finished). The resulting charging energy was 5.3 meV and the full width at half maximum (FWHM) of the transmission function was 160 µeV. This QD is the same as QD1 in our previous study on the lineshape of the thermopower of QDs [18]. QD2 was a 15 nm InAs segment defined again by InP barriers embedded in a 55 nm diameter InAs nanowire grown by CBE. This resulted in a classical charging energy of 6 meV and an energy spacing due to quantum confinement effects of up to 25 meV. Owing to thicker InP barriers than those of QD1, the FWHM of the transmission function of QD2 can only be quantified as much less than $kT$ ($T$ = 590 mK). Finally, QD3 consisted of a 72 nm diameter InSb nanowire grown by metal-organic vapor phase epitaxy (MOVPE) [32]. In this case the potential barriers that formed the QD were created by the Schottky Ti/Au contacts also serving as source and drain leads [33]. The distance between the two contacts was about 140 nm resulting in a charging energy of 5 meV. The FWHM of the transmission function was on the order of 100 µeV.

All three devices were fabricated in a similar way. Nanowires from the growth substrates were deposited on pre-patterned Si/SiO$_2$ chips (which also served as global back-gates) and contacted (see figure 1a for the contact configuration used) via standard electron-beam lithography processes, with a sulfur passivation treatment before evaporation of the contact materials [34], namely Ni/Au in the case of QD1 and QD2, and Ti/Au in the case of QD3.

To create a $\Delta T$ across the QD, an AC heating current $I_H$ (ranging from 10 to 100 µA) is run through the source contact (indicated by the red arrow in figure 1a), raising the temperature of the electron gas in that contact via Joule heating. $I_H$ is the result of two AC heating voltages, $V_{H\pm}$, tuned to have equal amplitudes and a 180° relative phase difference so that they sum to zero at the nanowire, thus biasing the nanowire thermally but not electrically. The third contact lead seen in figure 1a assists in balancing $V_{H\pm}$. Heating in this way not only raises the electron temperature above $T_0$, the original electron temperature before heating, in the nanowire source contact, but also in the drain contact [35]. This phenomenon appears to be mediated by electron-phonon coupling in the nanowire [36]. Therefore, $\Delta T$ is the local temperature difference between either side of the QD, and both of these temperatures are elevated above $T_0$. For QD2, the resulting $\Delta T$, caused by $I_H$, was measured using QD thermometry [35] and for QD1 modeled using finite element modeling [35]. An estimate for the $\Delta T$ for QD3 is given in Section 5.

We measured the AC thermovoltage, $V_{th}$, and thermocurrent, $I_{th}$, to evaluate the TE behavior of the QDs using lock-in techniques. We used frequencies of 13 Hz (QD1 and QD3) and 62.5 Hz (QD2) for $I_H$ and measured the second harmonic of the device current $I_{th}$ or open-circuit voltage $V_{th}$. As the open-circuit $V_{th}$ develops in response to the applied $\Delta T$, equilibrium is established by letting the drain float freely, and $V_{th}$ was measured using a high-impedance (1 TΩ) low-noise voltage preamplifier [18]. When measuring $I_{th}$, the hot source electrochemical potential, $\mu_H$, was fixed by the voltage source while the drain electrochemical potential, $\mu_C$, was grounded via a low-noise current preamplifier.

### 3. Nonlinear thermovoltage

In figure 1 we show data that demonstrate strongly nonlinear behavior of a QD's thermovoltage, measured at an electron temperature of $T_0$ = 240 mK. To characterize the dot, we first show in figure 1b a conductance peak formed by Coulomb blockade [37], measured as a function of gate voltage. The corresponding $V_{th}$ for the same gate voltage range (figure 1d) shows the characteristic lineshape that resembles the energy-derivative of the conductance peak [17, 38], and which, for small $\Delta T$, can be understood in detail based on the lineshape of the transmission resonance (figure 1b) as described in Ref. [18] using data from the same dot. Note that for increasing $\Delta T$ (on the order of $\Delta T \approx T_0$, see figure 1d) the $V_{th}$ signal doesn't simply scale linearly with applied thermal bias: with increasing $\Delta T$, the $V_{th}$ signal changes its shape, and the maxima of $V_{th}$ shift position. As a result, $V_{th}$ measured at fixed $V_g$ as a

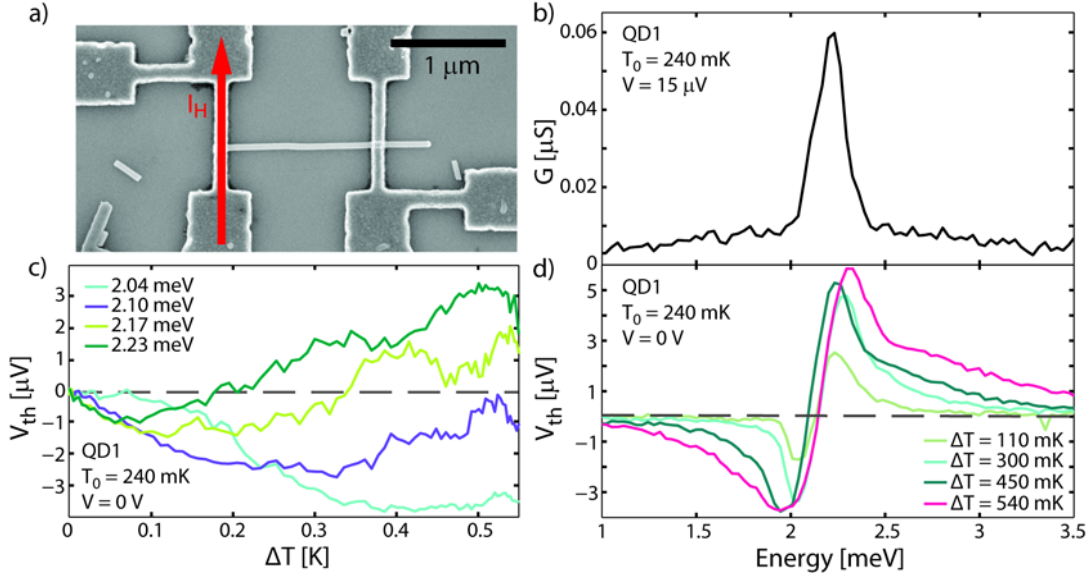

Figure 1: General device design and measured data from QD1.

a) Scanning electron microscopy (SEM) image showing the typical device design used for all devices in this study. For more details on the measurement configuration, see main text and Refs. [18, 36].
b) Differential conductance measured at $T_0 = 240$ mK as a function of gate voltage (converted to energy using the gate lever arm). For more details on this dot, see Ref. [18].
c) Thermovoltage measured as a function of $\Delta T$ at four different gate voltages (dot chemical potentials). The range of $\Delta T$ corresponds to a range of $I_H$ from 0 - 100 µA.
d) Thermovoltage measured at four different $\Delta T$ for the same gate voltage range as that used for the differential conductance data shown in b).

function of $\Delta T$ exhibits highly nonlinear behavior, including a sign reversal of $V_{th}$ (figure 1c). A similar reversal of a QD's $V_{th}$ as a function of $\Delta T$ was observed already in the first such experiments [17], but was at the time not understood.

To discuss phenomenologically the origin of nonlinear thermovoltage of the type shown in figure 1, we first consider the electrostatic back-action of the piled-up charges, which creates $V_{th}$ and shifts the position of the Coulomb blockade resonance. As a first approximation we neglect spin effects and consider electron tunneling between the leads and the QD as a sequence of single-electron transfers. For sufficiently large charging energies, transport can be described by switching between two charge states of the QD only. In this sequential tunneling limit, a master equation approach [39, 40] yields the current through the dot

$$I = \frac{2q\Gamma_L\Gamma_R}{h\Gamma}[f_H(\varepsilon_0) - f_C(\varepsilon_0)], \qquad (1)$$

where $q = -e$ is the electron charge, $f_H(\varepsilon_0) = 1/\{\exp((\varepsilon_0 - \mu_H)/[k(T + \Delta T)]) + 1\}$ and $f_C(\varepsilon_0) = 1/\{\exp((\varepsilon_0 - \mu_C)/[kT]) + 1\}$ are Fermi-Dirac distributions with electrochemical potentials $\mu_H = \mu^{eq} + qV_H$ and $\mu_C = \mu^{eq} + qV_C$ ; $\Gamma_L$ , $\Gamma_R$ are tunnel-induced broadenings ($\Gamma = \Gamma_L + \Gamma_R$); and $T + \Delta T, T$ are the temperatures on the left, L (hot, H) and the right, R (cold, C), sides respectively. Here $\varepsilon_0 = \varepsilon_0^{eq}(V_g) + \Delta\mu_{QD}(V)$ is the QD energy level with respect to the equilibrium chemical potential of the leads $\mu^{eq} = 0$. The equilibrium position $\varepsilon_0^{eq}(V_g)$ of the resonance is controlled by the gate voltage $V_g$ while the charge-polarization induced electrostatic shift of the QD potential $\Delta\mu_{QD}(V) = q(C_L V_H + C_R V_C)/C$ is determined by junction capacitances $C_L$ and $C_R$, with $C = C_L + C_R$.

We determine the thermovoltage $V_{\text{th}} = (\mu_H - \mu_C)/q$ from Eq. (1) by setting $I = 0$:

$$qV_{\text{th}} = -\frac{\varepsilon_0^{eq} C \Delta T}{CT + C_L \Delta T}. \qquad (2)$$

In the limit of small temperature differences, $\Delta T \ll T$, we recover the well-known result of a constant thermopower, $S = V_{th}/\Delta T = -\varepsilon_0^{eq}/qT$. In general, Eq. (2) establishes a nonlinear dependence of the thermovoltage with the thermal bias. However, it cannot explain the sign reversal as function of $\Delta T$ observed in the experiments.

Let us now consider a general current formula [41]

$$I = \frac{2q}{h} \int [f_H(E) - f_C(E)] \tau(E, V_g, V, T, \Delta T) dE, \qquad (3)$$

where $\tau$ is a generalized transmission probability which can become a complicated function in the presence of interactions. In the sequential limit, $\tau$ can be described with a delta-like function, $\tau \propto \delta(E - \varepsilon_0)$, and we recover Eq. (1). More generally, as indicated in Eq. (3), $\tau$ is a function of the electron energy $E$, the gate voltage $V_g$, the applied source-drain voltage $V$ (according to the aforementioned $\Delta\mu_{QD}$ generated by the capacitive dot-lead coupling [9, 10]), as well as on the temperatures in each lead, that is, on $T$ and $\Delta T$ [23]. Each of these dependencies can in principle lead to nonlinear TE effects [23, 24]: first, the part of $\tau(E)$ that is sampled by the electrons depends on $T$ and $\Delta T$, which can lead to a nonlinear TE response if $\tau(E)$ varies with $E$ on the energy scale $k\Delta T$; second, a self-consistent dependence of the dot's scattering potential (and thus of $\tau$) on $V$, $T$ and $\Delta T$ also leads to nonlinear TE behavior because the dot's scattering potential changes as $T$ is increased and a thermovoltage develops.

In the following, we first consider the effect of the energy dependence of $\tau(E)$ on nonlinear thermocurrent (Section 4), before turning to the influence of $T$ and $\Delta T$ (Section 5).

## 4. Energy-dependent transmission function

We begin by assuming $\tau$ to be independent of $T$ and $\Delta T$ and turn to measurements of the thermocurrent (rather than the thermovoltage), because the bias voltage, $V = (\mu_H - \mu_C)/q$, is constant when measuring $I_{\text{th}}$ (unlike when measuring $V_{\text{th}}$). We further assume that a constant $V$ just shifts $\tau(E, V)$ in energy by $\Delta\mu_{QD}(V)$, implying that the energy-dependent transmission can be described by a single-argument function which for symmetric couplings ($C_L = C_R$) takes the form $\tau(E, V) = \tau(E - qV/2)$. Furthermore, a large DC bias applied while measuring $I_{th}$ separates $\mu_H$ and $\mu_C$ energetically and thereby isolates the influence of the source and drain Fermi-Dirac distributions. For these reasons, $I_{th}$ is more straightforward to model than $V_{th}$.

Our strategy is to use current measurements at low temperature to obtain a good estimate of $\tau(E)$ for QD2, and to then compare the calculated, nonlinear thermocurrent based on this measured $\tau(E)$ with the experimental thermocurrent. During current measurements, the drain is grounded so that $\mu_C = 0$ while the source is biased so that $\mu_H = qV$. First, we measure the DC current, $I$, as a function of $V$ in a range where the current exhibits a simple, step-like behavior as is expected for single-level-dominated

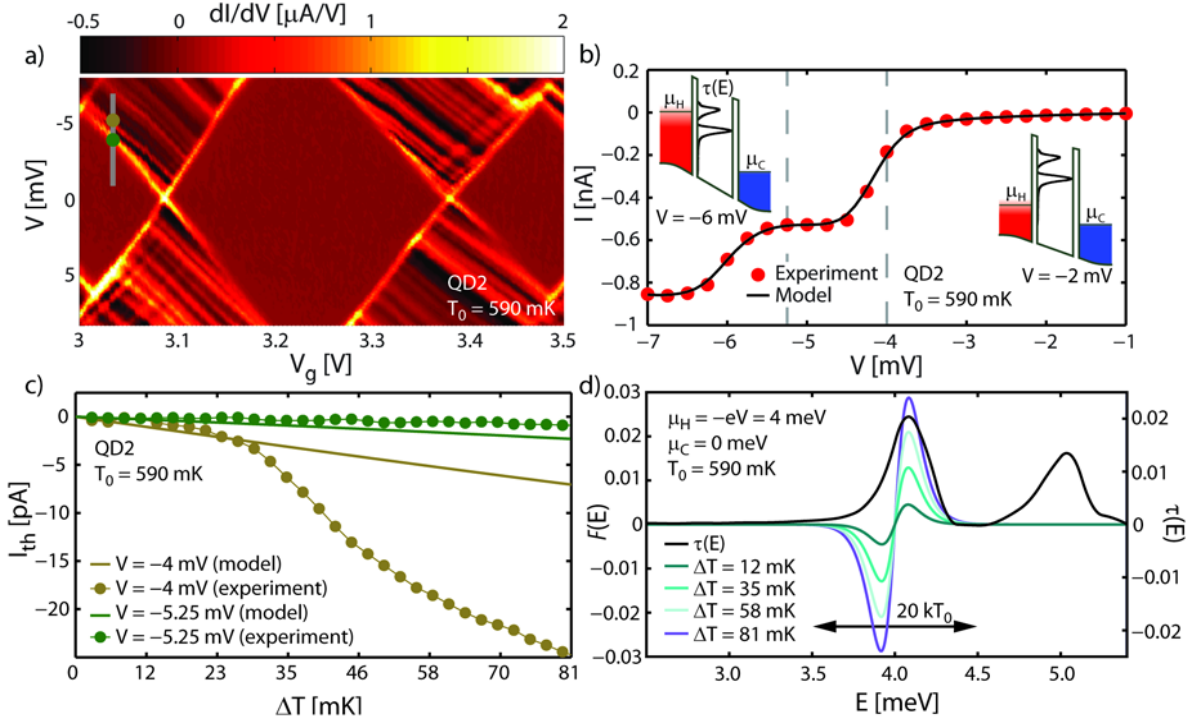

Figure 2: Nonlinear thermocurrent and model for QD2.

a) Coulomb blockade diamond for QD2 at $T_0 = 590$ mK. The grey line represents where the data in b) was measured. The two dots indicate where the data in c) was measured.

b) Measured DC current (dots) as a function of bias voltage, $V$. This data was used to deduce $\tau(E)$ shown in d) (see main text for details). Using $\tau(E)$, the numerically calculated $IV$ curve (black line) faithfully reproduces the data. The vertical dashed lines indictate where the data in c) was measured.
Insets: Schematic illustrations of the dot's band structure at a large $V$ when $\mu_C$ is situated far away from the resonance level of the quantum dot (the condition under which the data in c) were obtained).

c) The dotted lines show experimental data measured at $T_0 = 590$ mK as a function of $\Delta T$ at two different $V$. The theoretical predictions (solid lines) do not reproduce the strong nonlinear behavior seen in the measurements.

d) The transmission function $\tau(E)$ extracted from the measured data in b) is shown in black. The other curves show calculated $F(E) = f_H(E) - f_H^{\Delta T=0}(E)$, for four different $\Delta T$.

tunneling through a QD (see figure 2b for the current; the location of the sweep in the dot's stability diagram is shown in figure 2a). Next we use the relation $\tau \simeq G/(2q^2/h)$, which follows from Eq. (3) if $\tau$ is significantly wider than the thermal broadening of the source and drain Fermi edges, to extract $\tau(E,V)$ from the differential conductance $G = dI/dV$. The resulting $\tau(E)$ is shown in figure 2d. In figure 2b, we compare the original $IV$ data to the $IV$ curve calculated using Eq. (3) and the extracted $\tau(E,V)$ and find excellent agreement, confirming our assumption of an intrinsically broadened resonance. Equipped with $\tau(E,V)$, we turn our attention to the nonlinear thermocurrent.

The thermocurrent (measured at fixed $V$ and at the same $V_g$ as used in figure 2b and shown as dotted lines in figure 2c) is clearly nonlinear in $\Delta T$. We wish to determine whether this nonlinearity can be explained based on $\tau(E,V)$ alone by calculating the expected $I_{th}$. The data in figure 2c were measured at large negative $V$ ($|eV| \gg kT_0 \simeq 47$ µeV) to minimize the overlap between $\tau$ and $f_C$ (for a schematic illustration see insets of figure 2b). Therefore, $f_C(E)\tau(E,V) \simeq 0$, and $f_C$ can be dropped from Eq. (3). The thermocurrent through the QD can then be written as

$$I_{th} = I(\Delta T) - I(\Delta T = 0) = \frac{2q}{h} \int F(E)\,\tau(E,V)dE, \qquad (4)$$

where $F(E) = f_H(E) - f_H^{\Delta T=0}(E)$.

Examples of the calculated $I_{th}$ are weakly nonlinear in $\Delta T$ (solid lines in figure 2c). To understand the origin of this nonlinearity we plot the transmission function together with $F(E)$ for different $\Delta T$ (figure 2d) at $V = -4$ mV. For small $\Delta T$, electron and hole transport takes place in only a narrow region around $\mu_H$, and $\tau(E,V)$ for such a narrow region is comparable for electrons and holes. The resulting net $I_{th}$ is therefore small. As $\Delta T$ is increased, the energy range in which transport takes place gets broader, and the asymmetry of $\tau(E,V)$ around $\mu_H$ changes. Specifically, electron transport (region of positive $F(E)$) is enhanced by larger $\Delta T$, whereas hole transport (region of negative $F(E)$) is limited by the vanishing $\tau(E,V)$. The thermocurrent therefore increases nonlinearly to larger negative values.

However, comparing the modeled $I_{th}$ (lines, figure 2c) to the measured data (dots) we find that the model based on Eq. (3) can reproduce the experiment only for very small $\Delta T$. Already at $\Delta T$ larger than a few 10 mK, the experiment shows stronger nonlinear effects. Crucially, our simple model took into account only the energy dependent $\tau(E)$. In the next sections we address the effect of $T$ and $\Delta T$ on $\tau$.

## 5. Temperature-dependent transmission function

We now show that a temperature-dependent transmission function can explain the nonlinear behavior seen in our TE measurements. To do this, we turn to QD3, which again shows clearly nonlinear $I_{th}$, (measured at V = 0 V and $V_g$ fixed) including an example of a sign reversal at a specific $V_g$ (figure 3a). To explore the effect of a dependence of the transmission function on thermal bias, $\tau(E, \Delta T)$, we also show a 3D plot of $I_{th}$ as a function of both the heating voltage $V_H$ and $V_g$ (figure 3b). A precise calibration of $\Delta T$ as a function of $V_H$ is not available for this experiment. Based on device geometry and behavior, we estimate that $\Delta T$ is roughly quadratic in $V_H$ and that the largest $\Delta T$ achieved was on the order of 100 mK. As a function of gate voltage, the thermocurrent shows a resonant lineshape (see inset of figure 3b for one example of $I_{th}(V_g)$ at fixed $V_H$) very similar to that expected for the thermovoltage (figure 1d). Importantly, however, we find that the position of the thermocurrent resonance shifts as a function of $V_H$. This is highlighted in figure 3b by the black line, which indicates the position where $I_{th}$ changes sign. This trend, which most likely traces an underlying shift of the Coulomb blockade peak as a function of the temperatures in the dot leads [39, 42], is clear evidence of a dependence of the transmission function on $T$, $\Delta T$, or both, and corresponds to a renormalization of the energy levels in the QD [23, 24].

We support our interpretation by returning to thermovoltage measurements for QD1 and Eq. (2). We must specify how the dot potential changes with the applied temperature difference. Quite generally, this requires a self-consistent calculation that is beyond the scope of the present work. It suffices for our purposes to consider in Eq. (2) a linear dependence of $\varepsilon_0$ with $\Delta T$, replacing $\varepsilon_0^{eq}$ with $\varepsilon_0^{eq} + zk\,\Delta T$ where z is a characteristic potential that measures the QD response to temperature shifts [23]. We here take z as a dimensionless fitting parameter. Clearly, for $z < 0$ the thermovoltage reverses its sign. The

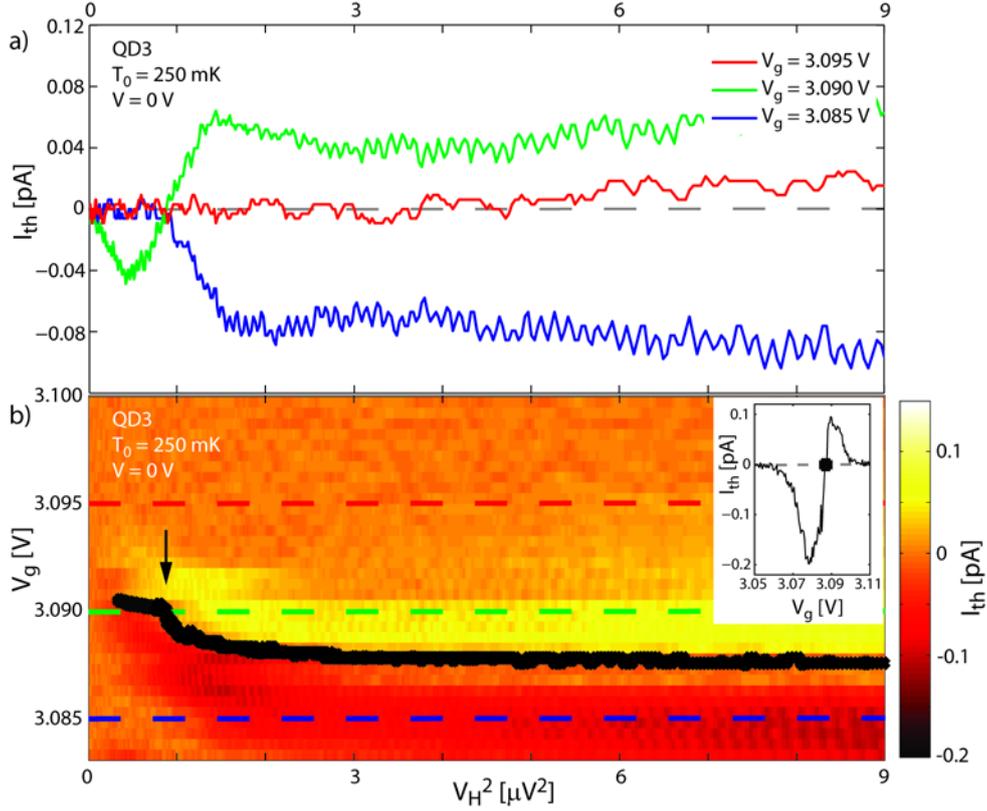

Figure 3: Nonlinear thermocurrent and energy level renormalization as a function of applied thermal bias for QD3.

a) Thermocurrent, $I_{th}$, as a function of squared heating voltage, $V_H^2$, measured at $T_0 = 250$ mK and $V = 0$ for three different $V_g$ as indicated.
b) A 3D measurement of $I_{th}$ as a function of $V_g$ and $V_H^2$ demonstrating that the QD energy level shifts as a function of $V_H$. The black arrow indicates where the sharpest shift (kink) occurs. This shift creates the strong nonlinear behavior seen in $I_{th}$. The black line is a guide to the eye showing where $I_{th}$ changes sign at resonance during each $I_{th}$ vs. $V_g$ trace (see inset for an example trace).
Inset: Example of $I_{th}$ as a function of $V_g$ at fixed $V_H$. The black dot contributes to the black line in (b).

precise value of the characteristic potential depends on the relative position of the QD level with respect to the leads' Fermi energy [23], thus we use different z for different $\varepsilon_0^{eq}$. In figure 4 we show a numerical calculation of $V_{th}$ as a function of $\Delta T$. The qualitative agreement with the experimental curves in figure 1c is fairly good. This demonstrates that the QD energy-level renormalization due to the applied thermal bias is crucial to understand the highly nonlinear dependence of $V_{th}$, and may be equally important for $I_{th}$. In Section 6 below, we argue that a possible microscopic source of such an effective level shift for QD3 may come from the thermally-induced change in the nature of transport excitations on the dot, e.g. break-down of spin-charge separation due to melting of residual Kondo correlations.

Based on these observations, the behavior of the thermocurrent as a function of $V_H^2$ in figure 3a can now be explained phenomenologically. In particular, the reason for the observed sign reversal of $I_{th}$ can now easily be understood by looking at the green dashed line in figure 3b: when measuring the thermocurrent at fixed $V_g$, the position of the energy level can shift past this $V_g$ as a function of $\Delta T$, causing the sign change observed in figure 3a. In a similar manner, one can track and fully explain the nonlinear behavior of the other two traces shown in figure 3a (see dashed lines in the corresponding colors in figure 3b).

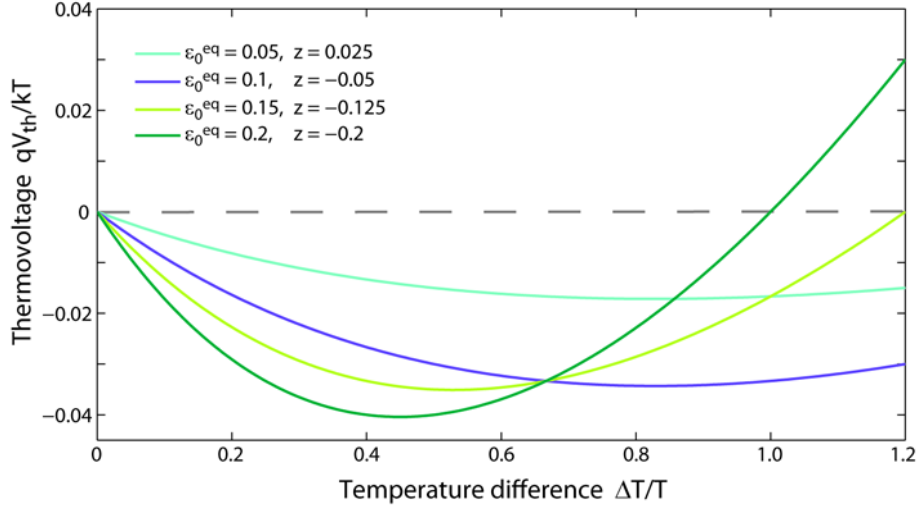

Figure 4: Thermovoltage obtained from a sequential-tunneling model as a function of the applied thermal bias for different values of the equilibrium level position $\varepsilon_0^{eq}$ (in units of $kT$) and the characteristic potential z.

## 6. Thermocurrent resonance as a probe of spin-charge separation transition

Our discussion in Section 5 showed that an energy-level renormalization as function of $\Delta T$ can explain the observed nonlinear TE behavior, but raises another question: what mechanism is responsible for the level shift? A particularly tantalizing observation is a sharp feature in the effective energy level position as a function of applied thermal bias (marked by an arrow in figure 3b) at around $V_H^2 = 1\,\mu V^2$. Simple mechanisms (such as asymmetric coupling to the leads, density-of-states modulation in the leads, or unequal bias drops across the barriers) cannot explain the severity of this kink, suggesting something more complex. This temperature-dependent kink might be indicative of a thermally induced phase transition (sharp crossover) either of the QD itself or of an external charge trap. As we argue below, one possible explanation is a correlations-driven spin-charge separation at low temperatures that alters the spectrum of the quasi-particle excitations on the QD that mediate thermotransport.

QD3 has been measured at zero voltage bias and sufficiently low base temperature for the Coulomb resonance to be dominated by a single orbital, which makes the single impurity Anderson model an appropriate framework to describe the thermotransport [20]. Many-body quantum correlations in tunneling-induced charge and spin fluctuations make the computation of the transmission (spectral) function $\tau(E)$ very challenging even in the linear response regime [43]. Nevertheless, we appeal to the well-known phase diagram of the Anderson model [44, 45] and make an approximate calculation [46] of the temperature-dependent energy level(-s) to support a microscopic interpretation of the strongly nonlinear evolution of the thermocurrent zero (see black line in figure 3b).

We assume that tuning $V_g$ past approximately 3.09 V gradually brings QD3 from the mixed valence (on-resonance) into the local-moment (singly occupied orbital) regime of the Anderson model. The latter is known for a spin-charge separation transition manifested in the spectral function at low temperatures [44] as follows: a narrow Kondo peak, due to lead-mediated virtual spin-flips, gets pinned to the Fermi energy ($\varepsilon_K \approx 0$) while the position $\varepsilon_h$ of a broader charged excitation – the holon – keeps following the gate-controlled orbital energy level $\varepsilon_0$. The Kondo effect is easily destroyed by heating and is very sensitive to $V_g$: for $\varepsilon_0 < 0$, the characteristic crossover temperature, $T_K$, dies off exponentially, $kT_K(\varepsilon_0) = \Gamma \exp(\pi \varepsilon_0 / 2\Gamma)$. Here $\Gamma$ is the energetic half-width of the empty orbital and $\varepsilon_0 = \varepsilon_0^{\text{bare}} + (\Gamma/\pi) \ln(\pi U / 4\Gamma) + \Gamma/\pi$ [45] already includes the temperature-independent renormalization due to hybridization with the leads which is controlled by the Hubbard charging energy $U \gg \Gamma, |\varepsilon_0^{\text{bare}}|$.[9] At sufficiently negative $\varepsilon_0$, the width of the holon peak is $2\Gamma$ because the holon is essentially a hole which can be filled by electrons of either spin orientation. At sufficiently low temperatures ($kT \ll \Gamma$), the merger of two separate peaks into a single excitation happens when $\varepsilon_0$ is near zero (mixed valence region, $|\varepsilon_0| \lesssim \Gamma$ [47]). We suggest that the kink in the data (in which temperature and $\varepsilon_0$ are changed, see figure 3b) results from elevating temperature beyond where this transition can occur.

For an approximate qualitative analysis, we assume that the resonance position(-s) $\varepsilon_0^*$ in the transmission function $\tau(E, T, \Delta T, V = 0)$ can be approximated using an equilibrium Green function for electrons on the dot at an effective average temperature $T_{\text{av}}$ which is a function of $T, \Delta T$ and scales approximately as $V_H^2$. We use the equations of motion method [46] with a high-order mean-field decoupling scheme [48]. Following Sec. V of [46], we use the denominator of the Green function to identify the energy renormalization equation [47],

$$\varepsilon_0^* = \varepsilon_0 + \frac{\Gamma}{\pi} \{ \text{Re}\, \Psi(1 - i\varepsilon_0^*/2\pi k T_{\text{av}}) + \ln(\Gamma/2\pi k T_{\text{av}}) \}, \qquad (5)$$

where $\Psi$ is the digamma function. Eq. (5) is consistent with analytic renormalization theory of Haldane [47] although the exact values of numerical factors of order one (cf. Eq. (5.23) of [45]) are beyond reach of simple analytical methods [46, 47]. At $kT_{\text{av}} < 0.143\, \Gamma$ multiple roots to Eq. (5) are possible for a limited range of $\varepsilon_0$, see insets in figure 5. The region of multiple solutions corresponds to partial spin-charge separation. We estimate the energies of the coexisting Kondo and holon excitations as $\varepsilon_K(\varepsilon_0, T_{\text{av}}) = \max \varepsilon_0^*(\varepsilon_0, T_{\text{av}})$ and $\varepsilon_h(\varepsilon_0, T_{\text{av}}) = \min \varepsilon_0^*(\varepsilon_0, T_{\text{av}})$, respectively. Note that for sufficiently negative $\varepsilon_0$, such that $\tilde{T}_K(\varepsilon_0) \lesssim T_{\text{av}}$, the Kondo resonance "melts" and the $\varepsilon_K(\varepsilon_0) \approx 0$ branch comes to an end (see inset a of figure 5).[10] Similarly, as $\varepsilon_0$ is increased above 0, $\varepsilon_K(\varepsilon_0)$ merges with the single-root branch $\varepsilon_0^*(\varepsilon_0)$ as the holon ceases to be relevant (inset a of figure 5).

Depending on the relative spectral weights of the dressed excitations and on the energy window $k\Delta T$ probed by the thermal transport measurements, the sign-reversal of $I_{th}$ can be expected at some $\varepsilon_0 = \varepsilon_0^{\text{zero}}$ such that either the Fermi level $\mu^{eq} = 0$ is between $\varepsilon_K$ and $\varepsilon_h$, or, if the solution to Eq. (5) is single-valued, the level is at resonance, $\varepsilon_0^*(\varepsilon_0) = 0$. These conditions define the domain of possible $\varepsilon_0^{\text{zero}}(T_{\text{av}})$ including an area of spin-charge separation at $T_{\text{av}} < T^*$, see the shaded region in figure 5.

Since the characteristic width of the Kondo peak $kT_K(\varepsilon_0)$ rapidly becomes very narrow compared to the roughly constant width $2\Gamma$ of the holon excitation as $\varepsilon_0$ becomes more negative, we expect the actual $\varepsilon_0^{\text{zero}}(T_{\text{av}})$ to be closer to the lower (smallest $|\varepsilon_0|$) branch of the coexistence domain in figure 5, with a rather sharp change of slope around $T^*$, similar to the measured kink in figure 3b.

We note that our prediction of a critical crossover point $(\varepsilon_0^{\text{zero}}, T^*)$ is consistent with a recent linear thermotransport study by Costi and Zlatić [43] of the strongly correlated regime of the Anderson model. From their phase diagram of the linear thermopower (figure 4 in [43]) we estimate $kT^* \sim 0.4\, \Gamma$ at $U/\Gamma = 8$ which is of the same order of magnitude as our $T^* \sim 0.11\, \Gamma$. The experimental estimates of $k\Delta T \sim 100$ mK $\sim 10\, \mu$eV and $\Gamma \sim 100\, \mu$eV are consistent with the theoretical expectations. Clearly, a more systematic experimental and theoretical study is necessary to confirm our conjectured observation of a spin-charge separation transition in a thermally-biased QD. Nonetheless, the ability of the Anderson-model-based theory to predict qualitatively the observed kink in the thermocurrent sign reversal is a strong indication that Kondo physics contributes to the nonlinear thermocurrent measured using QD3.

---

[9] We consider a single Coulomb-blockade resonance and neglect the double-occupancy spectral peak at large positive energies, focusing on the "infinite-U" limit of the Anderson model [44].

[10] The equations-of-motion method is known to overestimate $k\tilde{T}_K(\varepsilon_0) = \Gamma \exp(\pi \varepsilon_0/\Gamma) > kT_K(\varepsilon_0)$ [46].

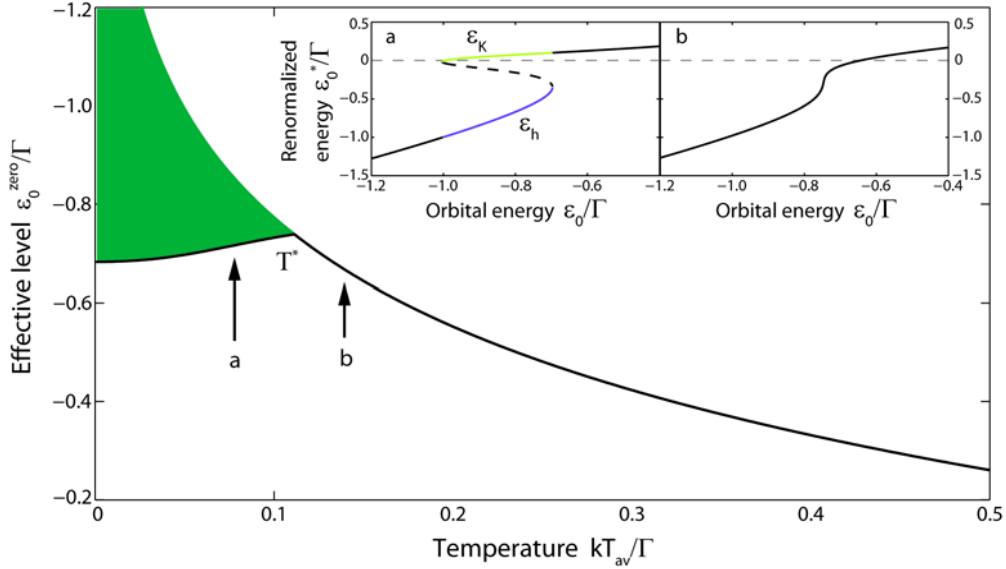

Figure 5: Estimation of the position of the effetive level $\varepsilon_0^{\mathrm{zero}}$ probed by thermocurrent based on the phase diagram of the large-U Anderson model. The shaded region marks the domain of partial spin-charge separation, as described in the text. The insets show solutions to Eq. (5) at two temperatures: $T_{\mathrm{av}} = 0.05\ \Gamma/k < T^*$ and $T_{\mathrm{av}} = 0.15\ \Gamma/k > T^*$.

## 7. Conclusion

We presented data from three quantum dots in two different material systems that each show strongly nonlinear TE behavior, indicating that the observed behavior is generic to quantum dots at low temperatures. We demonstrated that the energy dependence of the transmission function alone can account only for a very small part of the nonlinear behavior. Our analysis of the experimental data together with a phenomenological sequential-tunneling model demonstrate that most of the strongly nonlinear behavior appears to be related to heating-induced renormalization of the energy states in the QD, in agreement with recent predictions [23]. Furthermore, we propose a microscopic explanation of the observed renormalization in one of the dots (QD3) in terms of a spin-charge separation transition due to melting of residual Kondo correlations in the mixed-valence regime of the Anderson model.

Although low-dimensional systems have attracted the thermoelectrics community owing to possible efficiency gains via energy modulation and energy filtering, the associated nonlinear effects and underlying physical mechanisms had, prior to this work, not garnered much attention experimentally. The ability to control strongly nonlinear behavior in QDs, and the associated rectification effects, opens exciting avenues to explore a possible increase in the performance of TE energy converters [24]. In addition, measurements of the type shown in figure 3b (where energy level renormalization in a QD can be tracked in great detail using a heating current as a temperature control, revealing signatures of strong quantum correlations) may become a new, powerful tool to explore fundamental transport effects in QDs [20, 43]. In such future experiments, the effect of heating (an increase in average temperature $T$) and of a temperature difference ($\Delta T$) could be separated, for example, by defining two independent heaters at each end of the nanowire operated individually or in unison.

**Acknowledgements**

The authors thank Andreas Wacker and Olov Karlström for useful discussions, Ann Persson for access to her measurement data, Mingtang Deng for advice on sample fabrication, and Philippe Caroff for growth of the InSb nanowires. Financially supported by ONR, ONR Global, the Swedish Energy Agency (grant no. 32920-1), the Swedish Research Council (VR), the Thai government, NSF-IGERT, the Knut and Alice Wallenberg Foundation, the MINECO (grant no. FIS2011-23526), the Latvian Council of Science (grant no. 146/2012), the National Basic Research Program of the Ministry of Science and Technology of China (grant nos. 2012CB932703 and 2012CB932700), the National Natural Science Foundation of China (grant no. 91221202), and the Nanometer Structure Consortium at Lund University (nmC@LU).